\documentclass[11pt]{article}

\usepackage{amsmath}
\usepackage{graphicx}
\usepackage{amsfonts}
\usepackage{amssymb}
\usepackage{epsfig}
\usepackage{color}
\usepackage{psfrag}
\usepackage{epstopdf}

\newcommand{\EQ}{\begin{equation}}
\newcommand{\EN}{\end{equation}}
\newcommand{\be}{\begin{equation}}
\newcommand{\ee}{\end{equation}}
\newcommand{\bea}{\begin{eqnarray}}
\newcommand{\eea}{\end{eqnarray}}

\begin{document} \setcounter{page}{0}
\topmargin 0pt
\oddsidemargin 5mm
\renewcommand{\thefootnote}{\arabic{footnote}}
\newpage
\setcounter{page}{0}
\topmargin 0pt
\oddsidemargin 5mm
\renewcommand{\thefootnote}{\arabic{footnote}}
\newpage
\begin{titlepage}
\begin{flushright}
%SISSA 40/2012/EP \\
%DFTT 9/2007
\end{flushright}
\vspace{0.5cm}
\begin{center}
{\large {\bf Critical points of coupled vector-Ising systems. Exact results}}\\
%{\bf  of random cluster and $O(n)$ models}}\\
\vspace{1.8cm}
{\large Gesualdo Delfino$^{1,2}$ and Noel Lamsen$^{1,2}$}\\
\vspace{0.5cm}
{\em $^1$SISSA, Via Bonomea 265, 34136 Trieste, Italy}\\
{\em $^2$INFN sezione di Trieste, 34100 Trieste, Italy}\\
%{\em E-mail: delfino@sissa.it}\\
%\vspace{0.5cm}
%{\large and}\\
%\vspace{0.5cm}
%{\large P. Simonetti}\\
%\vspace{0.5cm}
%{\em Department of Physics, University of Wales Swansea,\\
%Singleton Park, Swansea SA2 8PP, United Kingdom}\\
%{\em email: p.simonetti@swansea.ac.uk}\\
\end{center}
\vspace{1.2cm}

\renewcommand{\thefootnote}{\arabic{footnote}}
\setcounter{footnote}{0}

\begin{abstract}
\noindent
We show that scale invariant scattering theory allows to exactly determine the critical points of two-dimensional systems with coupled $O(N)$ and Ising order pameters. The results are obtained for $N$ continuous and include criticality of loop gas type. In particular, for $N=1$ we exhibit three critical lines intersecting at the Berezinskii-Kosterlitz-Thouless transition point of the Gaussian model and related to the $Z_4$ symmetry of the isotropic Ashkin-Teller model. For $N=2$ we classify the critical points that can arise in the XY-Ising model and provide exact answers about the critical exponents of the fully frustrated XY model. 
\end{abstract}
\end{titlepage}

\newpage
When a statistical mechanical system possesses two order parameters, phase transitions associated with each of them can take place at different points of the phase diagram. It is possible, however, that the two types of ordering set in at the same point, and that this gives rise to novel critical behavior with new critical exponents. The example of a vector order parameter for $O(N)$ symmetry coupled to a scalar (Ising) order parameter for $Z_2$ symmetry is paradigmatic of the combination of continuous and discrete symmetries and was addressed since the early days of the perturbative expansion in $4-\varepsilon$ dimensions \cite{NKF}. The case $N=2$ (XY-Ising model) \cite{GKLN} has been higly debated in two dimensions also because it shares the ground state degeneracy of the fully frustrated (FF) XY model \cite{Villain} describing a Josephson-junction array in a magnetic field \cite{TJ}. The problem of whether this case can originate new critical behavior has been studied numerically for decades, with open questions persisting to this day  (see \cite{Kosterlitz} for a review). A consensus in favor of two transitions occurring at close but distinct temperatures very slowly emerged for the FFXY model (see \cite{SS,Olsson,HPV}), but disagreement on critical exponents remained even in the most extensive simulations (order $10^6$ lattice sites) \cite{HPV,OI,OYK}. For the XY-Ising model, which has a larger parameter space, two transition lines are observed to approach each other, without that the numerical analysis could so far determine the nature of the meeting point, although evidence for universal crossover effects in both models \cite{HPV} suggests the existence of a multicritical point with simultaneous criticality. The recent realization \cite{Struck} with cold atoms of a two-dimensional system with the symmetries of the XY-Ising model opened the way to experimental investigations of the critical behavior, but also here the required level of accuracy calls for theoretical benchmarking. On the analytic side, however, the problem has been considered as intractable, since the distance from the upper critical dimension as well as the interplay with the Berezinskii-Kosterlitz-Thouless (BKT) physics do not provide small expansion parameters, while an exactly solvable lattice realization of the coupled symmetries has never been found (see \cite{KNKB}). 

In this paper we show that the critical points of coupled $O(N)$ and Ising order parameters in two dimensions can be determined in a general and exact way, directly in the continuum limit. We obtain this result within the framework of scale invariant scattering \cite{paraf} that allowed, in particular, to progress with another longstanding problem such as critical quenched disorder \cite{random_mixed,DL2}. We determine the lines of renormalization group (RG) fixed points as a function of the variable $N$, which can be taken continuous, within a space of universal parameters. In particular, our results for $N=2$ allow us to classify the multicritical points that can arise in the XY-Ising model, and to draw conclusions about the critical exponents in the FFXY model. 

\begin{figure}
\begin{center}
\includegraphics[width=16cm]{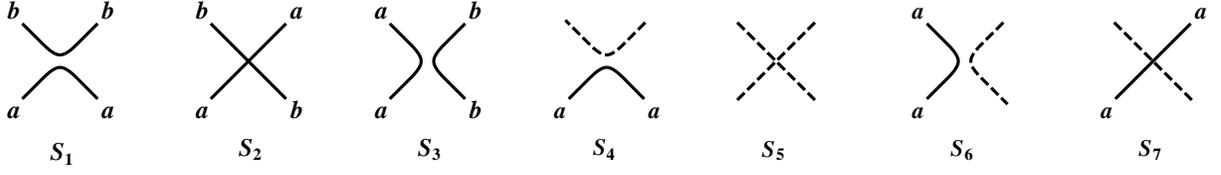}
\caption{Scattering processes for a vector particle multiplet ($a=1,2,\ldots,N$) and a scalar (dashed trajectories) at criticality. The amplitudes $S_1,\ldots,S_7$ are invariant under time (up-down) and space (right-left) reflections. 
}
\label{ON_ising_ampl}
\end{center} 
\end{figure}

We consider the two-dimensional vector-Ising model with lattice Hamiltonian
\EQ
{\cal H}=-\sum_{\langle i,j\rangle}[(A+B\sigma_i\sigma_j){\bf s}_i\cdot{\bf s}_j+C\sigma_i\sigma_j]\,,
\label{lattice}
\EN
which is invariant under the rotations of the $N$-component unit vectors ${\bf s}_i$ and the reversal of $\sigma_i=\pm 1$; the sum is taken over nearest-neighbor sites. We look for the points of simultaneous $O(N)$ and $Z_2$ criticality, where the correlators $\langle{\bf s}_i\cdot{\bf s}_j\rangle$ and $\langle{\sigma}_i{\sigma}_j\rangle$ behave as $|i-j|^{-2X_s}$ and $|i-j|^{-2X_\sigma}$, respectively, $X_s$ and $X_\sigma$ being the scaling dimensions; such points are fixed points of the RG (see e.g. \cite{Cardy_book}) where scale invariance allows to adopt the continuum description corresponding to a Euclidean field theory. We exploit the fact that such a theory is the continuation to imaginary time of a ($1+1$)-dimensional relativistic quantum field theory, which admits a description in terms of massless particles corresponding to the collective excitation modes. The combination of relativistic and scale invariance actually leads to conformal invariance \cite{DfMS}, which in two dimensions has infinitely many generators. These yield infinitely many conserved quantities, and then completely elastic particle scattering: the final state is kinematically identical to the initial one \cite{paraf}. This is why the two-particle scattering processes allowed at the RG fixed points are those depicted in Fig.~\ref{ON_ising_ampl}. The vector degrees of freedom correspond to a multiplet of particles $a=1,2,\ldots,N$, while the scalar corresponds to a particle whose trajectories we represent by dashed lines. The amplitudes $S_1,\ldots,S_7$ are those allowed by the requirement that the tensor, vector or scalar character of the initial state is preserved in the final state. 

Since the only relativistic invariant -- the center of mass energy -- is dimensionful, scale invariance leads to constant amplitudes \cite{paraf,fpu}. Then crossing symmetry \cite{ELOP}, which relates the amplitudes under exchange of space and time directions, takes the simple form
\bea
S_1=S_3^{*} &\equiv &  \rho_{1}\,e^{i\phi}, \\
S_2 = S_2^* &\equiv & \rho_2,\\
S_4 = S_6^* &\equiv &  \rho_4\, e^{i\theta}, \\
S_5 = S_5^*&\equiv & \rho_5, \\
S_7 = S_7^* &\equiv & \rho_7,
\eea 
and allows the parametrization in terms of the variables $\rho_1$ and $\rho_4$ non-negative, and $\rho_2$, $\rho_5$, $\rho_7$, $\phi$ and $\theta$ real. Finally, the unitarity of the scattering operator $S$ translates into the equations
\bea
 & \rho_1^2+\rho_2^2  =1,  \label{u1}\\
& \rho_1 \rho_2 \cos\phi =0,\label{u2}  \\
& N \rho_1^2 + \rho_4^2  +2\rho_1^2\cos2\phi =0, \label{u3} \\
 & \rho_4^2 + \rho_7^{2} = 1,\label{u4}\\
&  N\rho_4^2 + \rho_5^{2} =1,\label{u5}
\eea
\bea
&  \rho_4 \rho_7 \cos\theta = 0, \label{u6}\\
&\rho_4\left[\rho_2 e^{-i\theta}+\rho_1e^{-i(\phi+\theta)}+N\rho_1e^{i(\phi-\theta)}+\rho_5e^{i\theta}\right]=0. \label{u7}
\eea
For example, (\ref{u5}) follows from $1=\langle \o \o|SS^\dagger|\o \o\rangle=\langle \o \o|S\left[\sum_a|aa\rangle\langle aa|+|\o \o\rangle\langle \o \o|\right]S^\dagger|\o \o\rangle=N|S_4|^2+|S_5|^2$, where we denoted by $\o$ the scalar particle. Notice that $N$ enters the equations as a parameter that can be given real values. Such analytic continuation is well known for the decoupled $O(N)$ model, where it allows to describe self-avoiding walks for $N\to 0$ \cite{DeGennes}. 

\begin{table}
%[htbp]
%\centering
\begin{center}
\begin{tabular}{l|c|c|c|c|c|c}
\hline 
\hspace{-.5cm}Solution & $N$ & $\rho_2$ & $\cos\phi$ & $\rho_4$ & $\cos\theta$ & $\rho_5$ \\ 
\hline \hline
$\text{D}1_{\pm}$ & $\mathbb{R}$ & $\pm 1$ & - & 0 & - & $(\pm)1$ \\ 
$\text{D}2_{\pm}$ & $[-2, 2]$ & $0$ & $\pm\frac{1}{2}\sqrt{2-N}$ & 0 & - & $(\pm)1$ \\ 
$\text{D}3_{\pm}$ & $2$ & $\pm\sqrt{1-\rho_1^2}$ & $0$ & 0 & - & $(\pm) 1$  \\[0.3em] 
\hline 
$\text{F}1$ & $1$ & 0 & $[-\frac{1}{2}, \frac{1}{2}]$ & $\sqrt{1-4\cos^2 \phi}$ & 0 & $2\cos \phi$\\
$\text{F}2$ & $1$ & $[-1,1]$ & 0 & $\sqrt{1-\rho_2^2}$ & 0 & $\rho_2$\\
$\text{F}3$ & 1 & 0 & 0 & 1 & $[-1, 1]$ & 0 \\[0.3em]
\hline
$\text{L}1_{\pm}$ & $[-3, 1]$ & 0 & $\pm \frac{1}{2} \sqrt{1-N}$ & $1$ & $ (\pm)\frac{1}{2} \sqrt{1 - N} $ & $\pm \sqrt{1 -N}$ \\
$\text{L}2_{\pm}$ & $[-3, 1]$ & 0 & $\pm \frac{1}{2} \sqrt{1-N}$ & $1$ & $ (\pm) \frac{1}{2}\sqrt{3 + N}$ & $\mp \sqrt{1 - N}$  \\[0.3em]
\hline 
$\text{T}1_{\pm}$ & $(-\infty , 1]$ & $\pm \sqrt{\frac{1-N}{2-N}}$ & 0 & 1 & $ \frac{(\pm)1}{\sqrt{2}} \sqrt{1 (\pm) \frac{1}{\sqrt{2 - N}}}$ & $(\pm) \sqrt{1-N}$   \\[0.7em] 
$\text{T}2_{\pm}$ & $[-3, -2]$ & $0$ & $\pm 1$ & $ \displaystyle \sqrt{-2 - N } $ & 0 & $ \pm(N+1)$ \\ 
\hline 
\end{tabular} 
%}
%\vspace{5pt}
\caption{Solutions of the Eqs.~(\ref{u1})-(\ref{u7}) classifying the RG fixed points of two-dimensional systems with coupled $O(N)$ and Ising order parameters. One also has $\rho_1=\sqrt{1-\rho_2^2}$ and $\rho_7=(\pm)\sqrt{1-\rho_4^2}$. Signs in parenthesis are both allowed.
}
\label{solutions}
\end{center}
\end{table}

\begin{figure}
\begin{center}
\includegraphics[width=9cm]{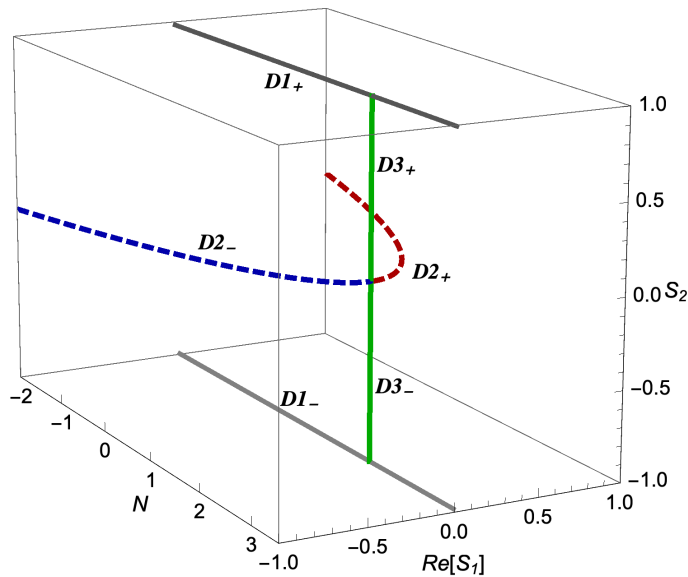}
\caption{Solutions of type D in the parameter space of the vector sector. This is decoupled from the scalar sector and describes, in particular, the critical lines of the dilute (D2$_-$) and dense (D2$_+$) regimes of the gas of nonintersecting loops, the BKT phase of the $XY$ model ($D3_+$), and the zero-temperature critical point of the $O(N>2)$ ferromagnet ($D1_+$).
}
\label{pure_space}
\end{center} 
\end{figure}

The solutions of the Eqs.~(\ref{u1})-(\ref{u7}) are given in Table~\ref{solutions} and provide the general and exact classification of the RG fixed points that can arise in the theory (\ref{lattice}). Their discussion conveniently begins with the solutions of type D (Fig.~\ref{pure_space}). These are characterized by $\rho_4=0$, amounting to decoupling between the vector and the scalar. Indeed, this yields $S_4=S_6=0$ and, recalling also (\ref{u4}) and (\ref{u5}), $S_5=S_7=\pm 1$. We recall that scattering in $1+1$ dimensions involves position exchange, and mixes statistics with interaction. It follows that $S_5=-1$ corresponds for the decoupled scalar sector to Ising criticality, which in two dimensions is described by a neutral free fermion \cite{DfMS}; on the other hand, $S_5=1$ accounts for the trivial fixed point (free boson). D1$_\pm$ corresponds to free bosons/fermions. In particular, the vector part (amplitudes $S_1,S_2,S_3$) of D1$_+$ also describes \cite{DL2} the asymptotically free zero-temperature critical point of the $O(N>2)$ ferromagnet (see e.g. \cite{Cardy_book}); hence, for $\rho_5=1$ the full solution D1$_+$ describes the zero-temperature critical point of the $O(N+1)$ model. The vector part of the solution D2$_\pm$ corresponds to nonintersecting trajectories ($S_2=0$) and was shown in \cite{paraf,DL2} to describe the critical lines of the gas of nonintersecting planar loops with fugacity $N$ (self-avoiding walks for $N\to 0$) in its dilute (D2$_-$) and dense (D2$_+$) regimes. This loop gas is known from its lattice solution (see \cite{Nienhuis}) to be critical for $N\in[-2,2]$; the correspondence between particle trajectories and loop paths was originally noted in the study of the off-critical case \cite{Zamo_SAW}. 

The solution D3$_\pm$ is defined for $N=2$ and contains $\rho_1$ as a free parameter. Its vector part then corresponds to the line of fixed points that accounts for the BKT transition \cite{BKT} in the XY model. We recall that this line is described by the Gaussian field theory with action 
\EQ
{\cal A}_{\text{Gauss}}=\frac{1}{4\pi}\int d^2x\,(\nabla\varphi)^2\,,
\label{gauss}
\EN
and energy density field $\varepsilon(x)=\cos 2b\varphi(x)$ with scaling dimension $X_\varepsilon=2b^2$; $b^2$ provides the coordinate along the line, with the BKT transition point corresponding to $b^2=1$, where $\varepsilon$ becomes marginal. Introducing the Euclidean complex coordinates $x_\pm=x_1\pm ix_2$, the equation of motion $\partial_+\partial_-\varphi=0$ yields the decomposition $\varphi(x)=\phi_+(x_+)+\phi_-(x_-)$. The fields 
\EQ
U_m(x)=e^{i\frac{m}{2b}[\phi_+(x_+)-\phi_-(x_-)]}\,,\hspace{1cm}m\in Z\,,
\label{vertex}
\EN
with scaling dimension $m^2/8b^2$, satisfy the condition that $\langle\cdots\varepsilon(x)U_m(0)\cdots\rangle$ is single valued in $x$ (see e.g. \cite{fpu}). $(\phi_+-\phi_-)/2b$ is the $O(2)$ angular variable, and the vector field ${\bf s}=(s_1,s_2)$ corresponds to $s_1\pm is_2=U_{\pm 1}$. The mapping on the solution D3$_\pm$ is provided by $\rho_1=\sin\frac{\pi}{2b^2}$ \cite{paraf,fpu}, so that the BKT phase corresponds to D3$_+$: it goes from the BKT transition point $b^2=1$ (contact point with D2$_\pm$ in Fig.~\ref{pure_space}) to the zero-temperature point $b^2=\infty$ (contact point with D1$_+$). 

\begin{figure}[t]
\centering
\includegraphics[width=8cm]{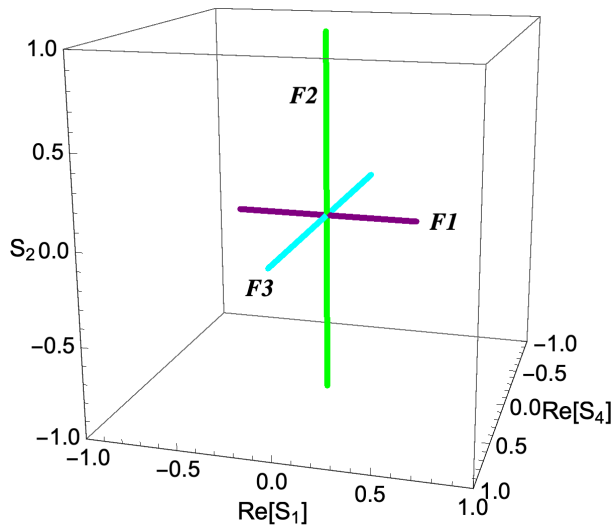}
\caption{Critical lines at $N=1$ due to the solutions of type F. They meet at the BKT transition point.}
\label{N1}
\end{figure}

With this understanding, we can continue the discussion of the results of Table~\ref{solutions}. The solutions of type F are all defined for $N=1$ and correspond to fixed points of two coupled Ising order parameters; the Hamiltonian (\ref{lattice}) becomes that of the Ashkin-Teller (AT) model \cite{AT}. The ``isotropic'' ($A=C$) AT model is known to possess a line of fixed points that also maps on the theory (\ref{gauss}), with continuously varying $X_\varepsilon=2b^2$, fixed $X_s=X_\sigma=1/8$, and $b^2$ nonuniversally related to the four-spin coupling $B$ \cite{Baxter,KB,DG}. In fact, all solutions of type F possess a free parameter and describe three critical lines sharing a common point (Fig.~\ref{N1}). The identification of this point with the BKT transition point $b^2=1$ follows from the observation that F2 has $S_1,S_2,S_3$ equal to $S_4,S_5,S_6$, respectively, so that it reconstructs the vector part of D3; then we know that $S_2=0$ corresponds to $b^2=1$. Further insight is obtained considering the theory with action
\EQ
{\cal A}_\text{Gauss}+\int d^2x\,\{\lambda\,\varepsilon(x)+\tilde{\lambda}\,[U_4(x)+U_{-4}(x)]\}\,.
\label{Z4}
\EN
Since we saw that $U_{\pm 1}$ define the components of a $O(2)$ vector, the terms $U_{\pm 4}$ in (\ref{Z4}) break $O(2)$ symmetry down to $Z_4$. The scaling dimensions that we specified above imply that at $b^2=1$ all the fields in the integral in (\ref{Z4}) are marginal. The RG equations around $b^2-1=\lambda=\tilde{\lambda}=0$ where studied at leading order in \cite{JKKN} and give three lines of fixed points: $\lambda=\tilde{\lambda}=0$ and $b^2=1$, $\lambda=\pm\tilde{\lambda}$. It was then conjectured in \cite{Kadanoff} that these lines may persist to all orders. Our exact result of Fig.~\ref{N1} shows that this is indeed the case. The isotropic AT model does possess $Z_4$ symmetry: for $A=C$ the Hamiltonian (\ref{lattice}), which contains the Ising variables $s$ and $\sigma$, is invariant under rotations of the vector $(s,\sigma)$ by angles multiples of $\pi/2$. The maximal value of $b^2$ realized in the square lattice AT model is $3/4$ \cite{KB}, and only the line with varying $b^2$ plays a role. 

\begin{figure}
\centering
\includegraphics[width=9.5cm]{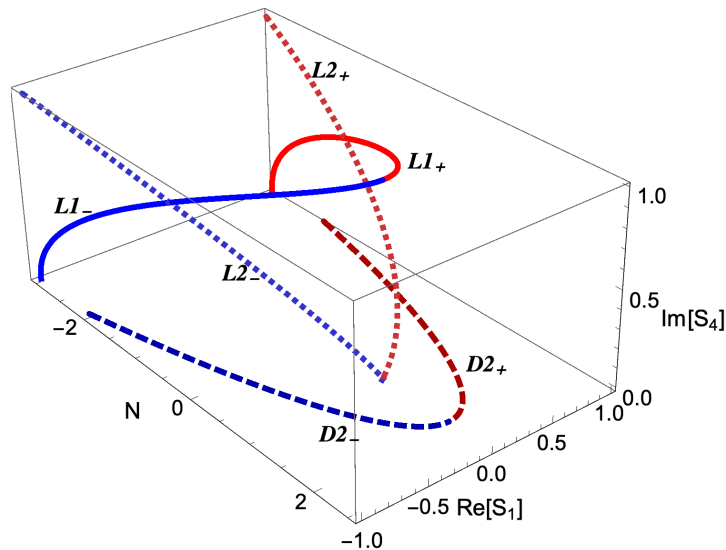}
\caption{The solutions L1 and L2 (continuous and dotted curves, respectively). As the vector part of D2 (dashed), they correspond to critical lines of the gas of nonintersecting loops.} 
\label{Ntype}
\end{figure}

The solutions of type L correspond to nonintersecting trajectories\footnote{Notice that $S_5$ cannot distinguish between intersection and nonintersection.} ($S_2=S_7=0$) and are defined for $N\in[-3,1]$ (Fig.~\ref{Ntype}). Hence, they reproduce the critical lines of the nonintersecting loop gas (vector part of solution D2) through a mechanism in which the scalar provides the second component of the vector. Finally, the solutions of type T cannot be traced back to the decoupled $O(N)$ case, and necessarily correspond to new universality classes. 

Let us now focus on the theory (\ref{lattice}) with $N=2$, i.e. on the XY-Ising model. We see from Table~\ref{solutions} that the only allowed RG fixed points are those of type D. We have already seen how the vector sector of D3 is related to the Gaussian theory (\ref{gauss}) and its parameter $b^2$; the scalar sector describes a trivial or Ising fixed point depending on the sign of $\rho_5$, with order parameter scaling dimension $X_\sigma$ equal to $0$ or $1/8$, respectively. The case $b^2=\infty$, $\rho_5=1$, with $X_s=X_\sigma=0$, describes the $O(3)$ zero-temperature critical point. Fig.~\ref{pure_space} shows that for $N=2$ the solution D3 includes as particular cases D1$_\pm$ ($b^2=\infty$ and $1/2$) and D2$_\pm$ ($b^2=1$).
%, with the caveat that the vector part of D1$_+$ also accounts for two decoupled trivial fixed points. 

Besides the points of simultaneous $O(2)$ and $Z_2$ criticality, which we can call multicritical, the XY-Ising model possesses points where only one of the order parameters is critical. Even considering those, it follows from our results that the only possible values for $X_\sigma$ at $N=2$ are 0 and $1/8$, while $X_s$ can vary continuously. However, since continuous symmetries do not break spontaneously in two dimensions \cite{nogold}, a vector ``ordering'' transition can only occur at $b^2=1$ through the BKT mechanism; hence, only the usual value $\eta_s=2X_s=1/4$ can arise at a vector transition point. 

There will be in the parameter space of the XY-Ising model phase transition lines bifurcating from a multicritical point and ending in an Ising critical point on one side and a BKT transition point on the other side. The $O(3)$ fixed point is a natural candidate for a zero-temperature multicritical point. The FFXY model can be defined on the square lattice through the Hamiltonian $-\sum_{\langle i,j\rangle}J_{i,j}{\bf s}_i\cdot{\bf s}_j$ ($J_{i,j}=\pm J$), with ferromagnetic horizontal rows and alternating ferromagnetic and antiferromagnetic columns. The model has the same ground state degeneracy of the XY-Ising model \cite{Villain}, but possesses only the parameter $J$. On universality grounds, it then corresponds to a line within the parameter space of the XY-Ising model. Our classification of allowed critical behaviors at $N=2$ implies that the exponents $0.2\lesssim\eta_\sigma=2X_\sigma\lesssim 0.4$, $0.8\lesssim\nu_\sigma\lesssim 1$ measured over the years (see the survey in \cite{HPV}) for the FFXY model are only consistent with the Ising universality class ($\eta_\sigma=1/4$, $\nu_\sigma=1$). Slow nonmonotonic approach to Ising exponents was observed in \cite{HPV} for increasing system size. We have also shown that at the vector transition only the BKT transition value $\eta_s=1/4$ is allowed. A check consistent with the BKT universality class was performed in \cite{HPV}, although $\eta_s$ was not measured. The value $\eta_s\simeq 0.2$ found in \cite{OYK} is instead not compatible with our results. Our conclusions on the FFXY exponents do not depend on simultaneous or separate transitions. The now accepted two-transition scenario suggests that the FFXY line intercepts the bifurcation originating from a multicritical point in XY-Ising parameter space. 

Summarizing, we have shown how scale invariant scattering theory yields the exact solution to the longstanding problem of determining the RG fixed points for two-dimensional systems with coupled $O(N)$ and Ising order parameters. For $N=2$ this enabled us to classify the multicritical points allowed in the XY-Ising model and to provide exact answers about the FFXY exponents. At $N=1$ we have exhibited three lines of fixed points intersecting at the BKT transition point of the Gaussian theory and related to the $Z_4$ symmetry of the isotropic AT model. For $N\leq 1$ new universality classes appear that can be relevant for gases of intersecting loops.

%\end{document}
%\newpage
%\vspace{1cm} \noindent \textbf{Acknowledgments.} 

\end{document}